\documentclass{article}

\usepackage{PRIMEarxiv}

\usepackage[utf8]{inputenc} 
\usepackage[T1]{fontenc}    
\usepackage{hyperref}       
\usepackage{url}            
\usepackage{booktabs}       
\usepackage{amsfonts}       
\usepackage{nicefrac}       
\usepackage{microtype}      
\usepackage{lipsum}
\usepackage{fancyhdr}       
\usepackage{graphicx}       

\usepackage{xcolor}
\usepackage{hyperref}
\usepackage{bbm}
\usepackage{amsmath}
\usepackage{pifont}
\usepackage{subcaption}
\usepackage{arydshln}
\usepackage{makecell}
\usepackage{multirow}
\title{VertMatch: A Semi-supervised Framework for Vertebral Structure Detection in 3D Ultrasound Volume}

\author{
  Hongye Zeng$^{1,2,3}$, kang Zhou$^{4}$, Songhan Ge$^{1,2,3}$, Yuchong Gao$^{1,2,3}$, \\
  \and \textbf{Jianhao Zhao$^{1}$, Shenghua Gao$^{1,5}$, Rui Zheng$^{1,5,}$\thanks{zhengrui@shanghaitech.edu.cn}} \\
  \\
  \small
  $^1$School of Information Science and Technology, ShanghaiTech University, Shanghai 201210, China\\
  \small
  $^2$Shanghai Advanced Research Institute, Chinese Academy of  Sciences, Shanghai, China\\
  \small
  $^3$University of Chinese Academy of Sciences, Beijing, China\\
  \small
  $^4$Department of Computer Science and Engineering, The Chinese University of Hong Kong, Hong Kong, China\\
  \small
  $^5$Shanghai Engineering Research Center of Intelligent Vision and Imaging, ShanghaiTech University, Shanghai, China\\
}

\begin{document}
\maketitle
\begin{abstract}
Three-dimensional (3D) ultrasound imaging technique has been applied for scoliosis assessment, but current assessment method only uses coronal projection image and cannot illustrate the 3D deformity and vertebra rotation. The vertebra detection is essential to reveal 3D spine information, but the detection task is challenging due to complex data and limited annotations. We propose VertMatch, a two-step framework to detect vertebral structures in 3D ultrasound volume by utilizing unlabeled data in semi-supervised manner. The first step is to detect the possible positions of structures on transverse slice globally, and then the local patches are cropped based on detected positions. The second step is to distinguish whether the patches contain real vertebral structures and screen the predicted positions from the first step. VertMatch develops three novel components for semi-supervised learning: for position detection in the first step, (1) anatomical prior is used to screen pseudo labels generated from confidence threshold method; (2) multi-slice consistency is used to utilize more unlabeled data by inputting multiple adjacent slices; (3) for patch identification in the second step, the categories are rebalanced in each batch to solve imbalance problem. Experimental results demonstrate that VertMatch can detect vertebra accurately in ultrasound volume and outperforms state-of-the-art methods. VertMatch is also validated in clinical application on forty ultrasound scans, and it can be a promising approach for 3D assessment of scoliosis.
\end{abstract}

\section{Introduction}
\label{sec1}
Adolescent idiopathic scoliosis (AIS) is a 3D deformity of spine featured with lateral deviation and axial vertebral rotation (\cite{weinsteinAdolescentIdiopathicScoliosis2008};\cite{asherAdolescentIdiopathicScoliosis2006}), and can greatly influence the patient’s life and even give rise to death (\cite{weinsteinHealthFunctionPatients2003}). Therefore, the early diagnosis and frequent follow-up exams of AIS are very important to the treatment. The radiography is usually used to monitor scoliosis (\cite{kimScoliosisImagingWhat2010}), but it can accumulate ionizing radiation for patients during the long-term follow-up exams (\cite{lawCumulativeRadiationExposure2016}). The 3D ultrasound imaging technique has become a promising diagnostic tool since it is radiation-free, cost-effective and portable comparing to radiography (\cite{huang2018};\cite{chenImprovement3DUltrasound2021}).
\begin{figure}[ht]
    \begin{minipage}[c]{0.3\linewidth}
        \centering
        \begin{minipage}[t]{0.97\linewidth}
            \centering
            \includegraphics[width=\linewidth]{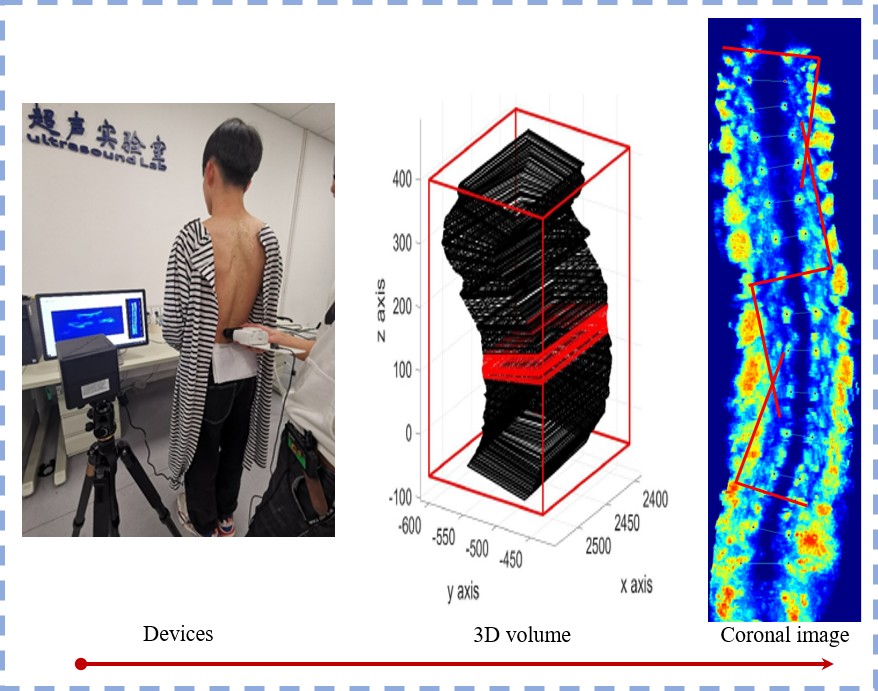}
            \subcaption{}
            \label{fig:intra_a}
        \end{minipage}
        \centering
        \begin{minipage}[t]{0.97\linewidth}
            \centering
            \includegraphics[width=\linewidth]{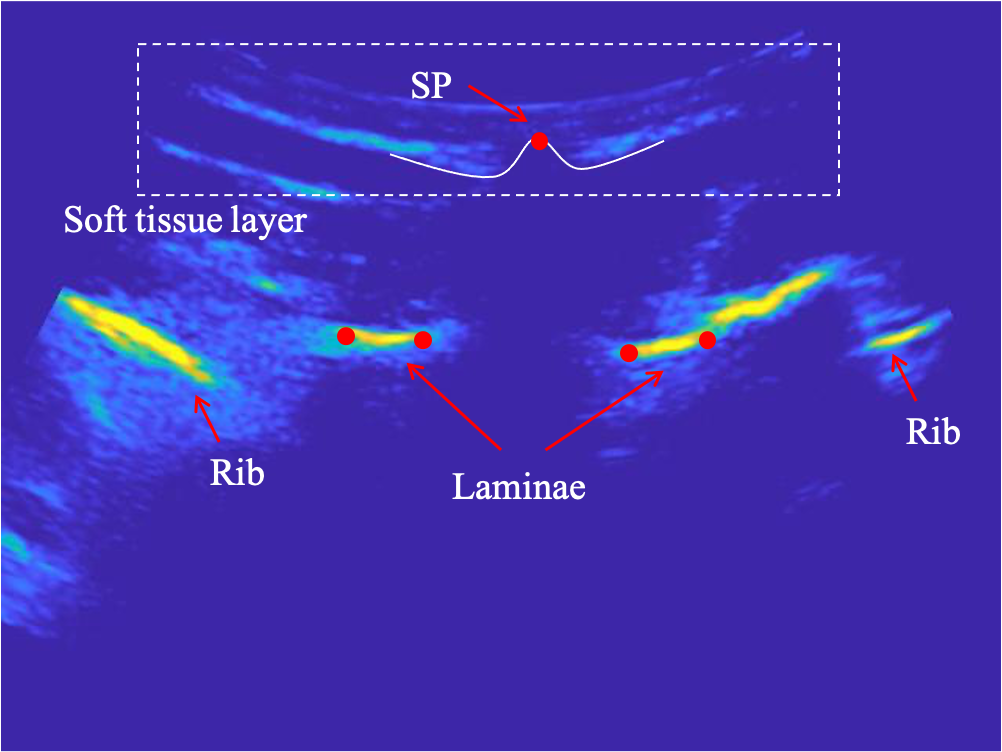}
            \subcaption{}
            \label{fig:intra_b}
        \end{minipage}
    \end{minipage}
    \centering
    \begin{minipage}[c]{0.6\linewidth}
        \centering
        \includegraphics[width=\linewidth]{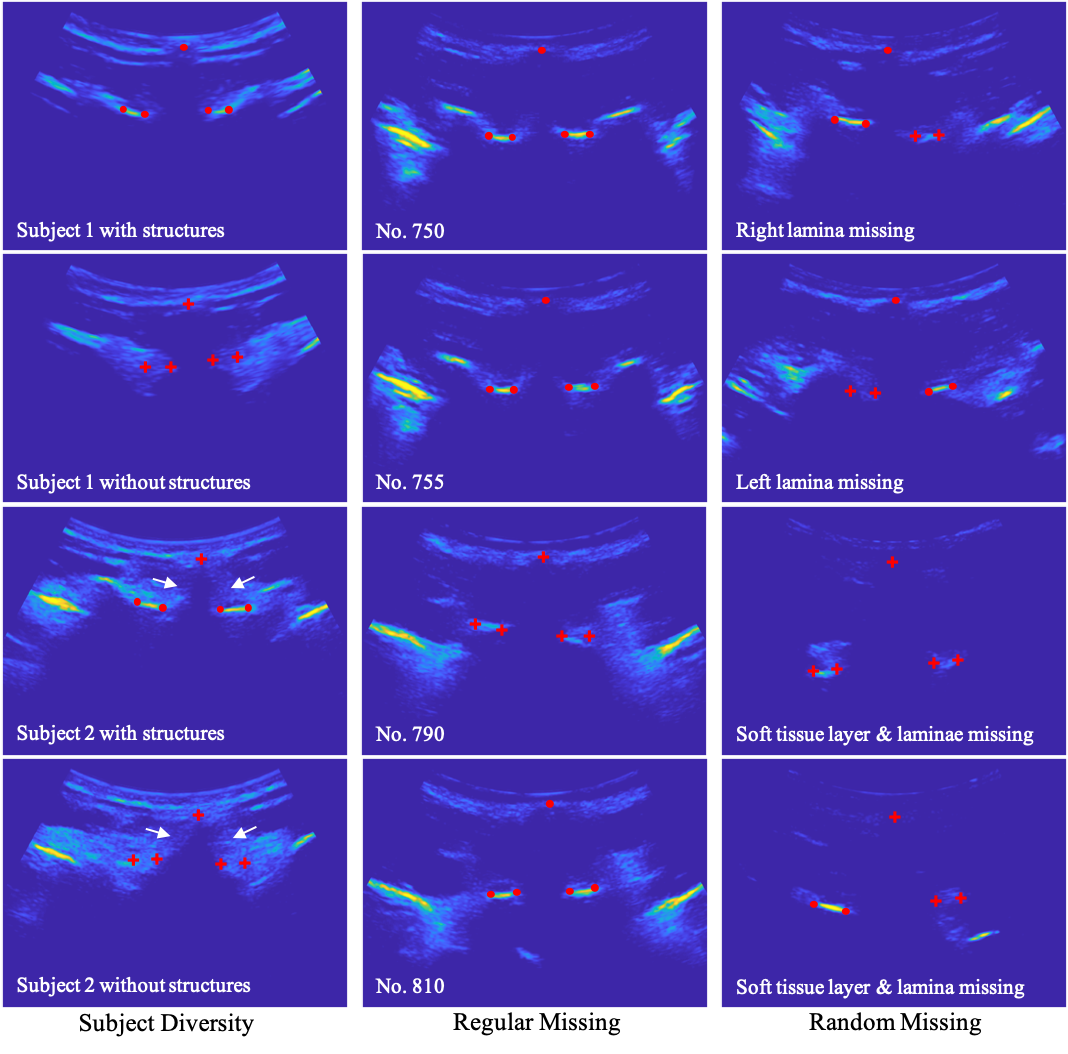}
        \subcaption{}
        \label{fig:intra_c}
    \end{minipage}
    \captionsetup{font={small}}
    \caption{Introduction of task and annotation: (a) AIS assessment only using coronal projection image of 3D ultrasound volume; (b) Vertebral structures on transverse slice. Five landmarks are annotated to represent possible positions of spinous process (SP) and laminae; (c) Challenging slices with annotation. These slice contain real (red dot) or fake (red cross) structures.}
    \label{fig:intro}
\end{figure}
As shown in Fig. \ref{fig:intra_a}, 3D ultrasound technique is usually used for AIS assessment by projecting the 3D volume to 2D coronal spine image and measuring the spinal curvature (\cite{chenReliabilityAssessingCoronal2013};\cite{cheungUltrasoundVolumeProjection2015};\cite{zhouAutomaticMeasurementSpine2017}). Although many researches have proved that coronal spine image is reliable for curvature measurement (\cite{zhengIntraInterraterReliability2015};\cite{zhengFactorsInfluencingSpinal2018}), the 2-D projection image cannot expose the 3D deformity and vertebra rotation. 

Therefore, it is necessary to further detect vertebra in the ultrasound volume and provide more 3D spine information for better AIS assessment. The main structures of vertebra to be detected are spinous process (SP) and laminae on transverse slice as shown in Fig. \ref{fig:intra_b}. The SP is usually merged into the soft tissue and presents an obvious hollow (white curve). The laminae are usually located in the middle region of the transverse slice and shown as the two adjacent short line segments with high intensity. However, it is challenging to detect SP and laminae on all transverse slices in volume due to three obstacles as illustrated in Fig. \ref{fig:intra_c}: 1) Subject Diversity: slices from two subjects especially noise (white arrow) are quite different, due to different fat and muscle content; 2) Regular Missing: from $750^{th}$ to $810^{th}$ slices from the same subject, the laminae and SP regularly disappears and reappears at the junction of two vertebrae; 3) Random Missing: the soft tissue layer or unilateral lamina disappear randomly, since the ultrasound signal may not be received by the probe or be blocked by other bones due to vertebra rotation or scanning manipulation. Subject diversity enriches the data and causes large demand of annotations for learning-based method, and missing problem requires the detection method to distinguish real vertebral structures accurately in complex ultrasound volume.

The straightforward thought is to apply existing vertebra detection method on 3D ultrasound volume. Some methods have been proposed on sagittal slices to segment the transverse process (\cite{ungiAutomaticSpineUltrasound2020}), which is quite different from SP and laminae on transverse slice. \cite{kamaliRealtimeTransverseProcess2018} and \cite{tangCNNbasedMethodReconstruct2021} successfully segmented the spinous process and spine surface on transverse slice respectively using the acoustic shadow method and U-Net. Excepting for the ultrasound image, many mature methods are applied to detect vertebrae on Computed Tomography (CT) (\cite{taoSpineTransformersVertebraLabeling2021}) and Magnetic resonance imaging (MRI) (\cite{zhanRobustMRSpine2012}). These methods usually uses the relation between several vertebrae as anatomical prior to achieve better predictions (\cite{wangAutomaticVertebraLocalization2021}), which is informative but not entirely suitable for our transverse slice containing information of single vertebra. Besides, above learning-based methods need large amounts of annotated data, which is very costly and time-consuming for medical experts especially on 3D images. The semi-supervised leaning (SSL) shows potential for improving network performance when labeled data is scarce by using pseudo-labeling and consistency regularization to utilize unlabeled data (\cite{xieSelfTrainingNoisyStudent2020};\cite{sohnFixMatchSimplifyingSemiSupervised2020}). Recently, the SSL algorithm has been successfully applied for general vision tasks (\cite{zhouInstantTeachingEndtoEndSemiSupervised2021};\cite{mittalSemiSupervisedSemanticSegmentation2021};\cite{sohnSimpleSemiSupervisedLearning2020}) and medical image analysis (\cite{zhouSSMDSemiSupervisedMedical2021};\cite{liuSemiSupervisedMedicalImage2020};\cite{huoAutomaticGradingAssessments2022};\cite{luoSemisupervisedMedicalImage2021}). They usually proposed some novelties of consistency regularization, which are instructive for our task.

In this paper, we propose a two-step landmark detection framework for vertebra detection in 3D ultrasound volume. We firstly consider the vertebral structure detection as landmark detection task to avoid complex segmentation annotations, which means that the SP and laminae are annotated as five landmarks as illustrated in Fig. \ref{fig:intra_b}. Specifically, five landmarks only represent possible positions of SP and laminae, and contain real (dots) and fake (cross) vertebral structures as shown in Fig. \ref{fig:intra_c}. Secondly, we design a two-step framework containing the detector to detect possible positions of SP and laminae globally, and the classifier to identify possible positions whether contain real structures by using local patch. On one hand, it is easier to detect possible position since five landmarks usually locate in the middle of slice and form an acute triangle. On the other hand, the fake structures are obviously different from real structures when focusing on local patch. In a word, two-step method can reduce difficulty of vertebral structure detection compared to find real structures in one step. 

To address the limited annotation problem, we propose to use SSL to train the detector and classifier, and propose three novel components for our task: 1) To generate more convinced pseudo labels, anatomical prior is leveraged to screen the pseudo labels produced from high confidence predictions. Anatomical prior is summarized that the SP and laminae usually form an acute triangle due to the vertebra shape, and the bone length usually is within an approximate range. 2) To utilize more unlabeled data, a slice and its adjacent slices are fed into the network together. This procedure is based on multi-slice consistency, which declares that there are only slight differences between adjacent slices while the possible positions of SP and lamina are supposed to be consistent. Therefore the detector can be trained by forcing to output similar predictions of adjacent slices. As regular missing slices shown in Fig. \ref{fig:intra_c}, although the SP and laminae disappear and reappear, the possible positions of SP and laminae are approximate on consecutive slices ($750^{th}$ and $755^{th}$) and locate at the center of slice. 3) To avoid classifier bias, the categories in pseudo labels are rebalanced in each batch, since the slices containing SP and laminae are minority in ultrasound scans.

The main contributions are summarised as follow:
\begin{itemize}
    \item We propose VertMatch framework for vertebral structure detection in 3D ultrasound volume. VertMatch contains two steps of position detection and patch identification, and can tackle complicated ultrasound slices from various subjects.
    \item We train VertMatch in semi-supervised manner and propose three novel components: for position detection, (1) anatomical prior is used to select better pseudo labels;  (2) multi-slice consistency is proposed to utilize more unlabeled data; (3) for patch identification, the batch rebalance is used to balance the categories in pseudo labels.
    \item We further verify VertMatch for AIS assessment on 40 subject scans by measuring spinous process angle (SPA), which is a clinical parameter to reveal axial vertebral rotation and lateral curvatures. The results of SPA from ultrasound data are comparable to the gold standard from radiograph, which shows VertMatch's potential for 3D AIS assessment.
\end{itemize}

\section{Related work}
\subsection{Vertebra detection}
The spine can be visualized using 3D imaging technique such as CT, MRI and ultrasound, and the vertebra localization and identification are required in many clinical applications of spine disorder diagnosis and surgery planning. 

Some methods have been explored for vertebra detection on ultrasound volume. \cite{bakaUltrasoundAidedVertebral2017}, \cite{kamaliRealtimeTransverseProcess2018}, \cite{tranAutomaticDetectionLumbar2010}, \cite{ungiAutomaticSpineUltrasound2020} proposed novel methods to segment transverse process on sagittal slices, and \cite{bertonSegmentationSpinousProcess2016}, \cite{tangCNNbasedMethodReconstruct2021} focused on spine surface segmentation on transverse slices. Most methods implemented their algorithm using the U-Net, and \cite{tangCNNbasedMethodReconstruct2021} proposed a late fusion U-Net to fuse the B-mode and shadow-enhanced B-mode image. All current methods for 3D ultrasound volume cannot tackle challenges in our task.  

Anatomical prior has been exploited for vertebra detection in CT and MRI volume, and these methods inspire us to derive and utilize new anatomical prior on ultrasound transverse slices. \cite{liaoJointVertebraeIdentification2018} used the Bi-RNN to incorporate the spine anatomical knowledge into neural networks implicitly, \cite{zhanRobustMRSpine2012} used a hierarchical learning strategy to detect different level of spine based on prior, and \cite{yangAutomaticVertebraLabeling2017} explicitly considered the spatial distribution of the vertebrae by using an information aggregation layer. More commonly, anatomical prior was used to post-process and constrain the network outputs, by using the Hidden Markov Model (\cite{chenVertebraeIdentificationLocalization2020}) or solving the anatomical-constraints optimization problem (\cite{wangAutomaticVertebraLocalization2021}).

\subsection{Landmark detection}
The vertebra detection on image volume described in 2.1 is usually implemented as a segmentation task which is time-consuming to annotate. While the vertebra detection on radiograph is usually formulated as a landmark detection task, since the landmark annotation is simpler and faster to accomplish compared to segmentation annotation. \cite{wuAutomaticLandmarkEstimation2017} used the BoostLayer to eliminate the outlier features for robust landmark detection. \cite{sunDirectEstimationSpinal2017} proposed the support vector regression, and \cite{zhangLearningbasedCoronalSpine2021} applied the HR-Net, to predict landmarks and estimated the spinal curvature directly. \cite{wuAutomatedComprehensiveAdolescent2018} proposed the multi-view correlation network to integrate the anterior-posterior and lateral radiographs, and achieved accurate results for landmark detection and curvature estimation. The landmark detection is a typical task for computer vision such as human pose estimation. The stacked hourglass network (SHN) (\cite{newellStackedHourglassNetworks2016}) is a commonly used network by encoding low-resolution representation and recovering high-resolution representation. In contrast, the high-resolution network (HRNet) (\cite{wangDeepHighResolutionRepresentation2021}) is another popular method and maintain high-resolution representations through the whole process.

\subsection{Semi-supervised learning}
The Semi-supervised Learning (SSL) has been extensively studied to efficiently utilize unlabeled images, and recent SSL methods usually use the pseudo-labeling and consistency regularization. Pseudo-labeling based method predicts pseudo labels for unlabeled data using a trained model, and then all data with real and pseudo labels are used to retrain the model. \cite{leePseudoLabelSimpleEfficient2013} was the first to introduce the pseudo-labeling for neural networks, and \cite{berthelotMixMatchHolisticApproach2019a} proposed MixMatch to average the predictions of stochastic augmented inputs by employing a sharpen function. Consistency regularization utilizes unlabeled data based on the assumption that the network should output similar predictions when being fed perturbed versions of the same image. \cite{laineTemporalEnsemblingSemiSupervised2017} proposed the temporal ensembling model to improve the quality of consistency targets by using the exponential moving average strategy for averaging predictions over the previous epoch, and \cite{tarvainenMeanTeachersAre2017} proposed the Mean Teacher framework which averaged model weights instead of label predictions by including an additional teacher model. FixMatch (\cite{sohnFixMatchSimplifyingSemiSupervised2020}) generated pseudo label on weakly-augmented unlabeled images, and the model was trained to predict the pseudo label when being fed a strongly-augmented version of the same image.

The SSL methods has been widely studied in medical image fields. \cite{wangFocalMixSemiSupervisedLearning2020} was first to apply the SSL for 3D medical image detection, and the SSL also were beneficial to the segmentation of muscle (\cite{pengCMCNet3DCalf2022}) and breast structures (\cite{liDeepWeaklysupervisedBreast2022}). Most of the methods proposed novel consistency regularization to suit their task. \cite{zhouSSMDSemiSupervisedMedical2021} proposed the adaptive consistency to regularize different components in the predictions. \cite{liuSemiSupervisedMedicalImage2020} proposed the sample relation consistency to model the relationship information among different samples. \cite{huoAutomaticGradingAssessments2022} used the attention consistency to ensure the focusing of the student and teacher networks on the same defect regions for cartilage defect level classification. \cite{luoSemisupervisedMedicalImage2021} defined the dual-task which consisted of pixel-level segmentation task and geometry-aware level set representation task, and proposed the dual task consistency which declared that the networks should behave similarly between two different tasks when being fed same image. In this study, we also propose new consistency regularization adapted to our task.

\section{Methods}
We propose a semi-supervised framework, termed as VertMatch, to detect vertebral structures in 3D ultrasound volume. To reduce task difficulty, we consider the vertebral structures detection as landmark detection task and annotate the possible positions of structures as five landmarks, furthermore we decompose entire task into two steps containing global detection and local identification. To efficiently use the data and avoid a large annotation, we train the networks using semi-supervised manner, and three novel components are proposed to improve SSL algorithm based on anatomical prior and observation of ultrasound data.

\begin{figure}[ht]
    \centering
    \includegraphics[scale=0.62]{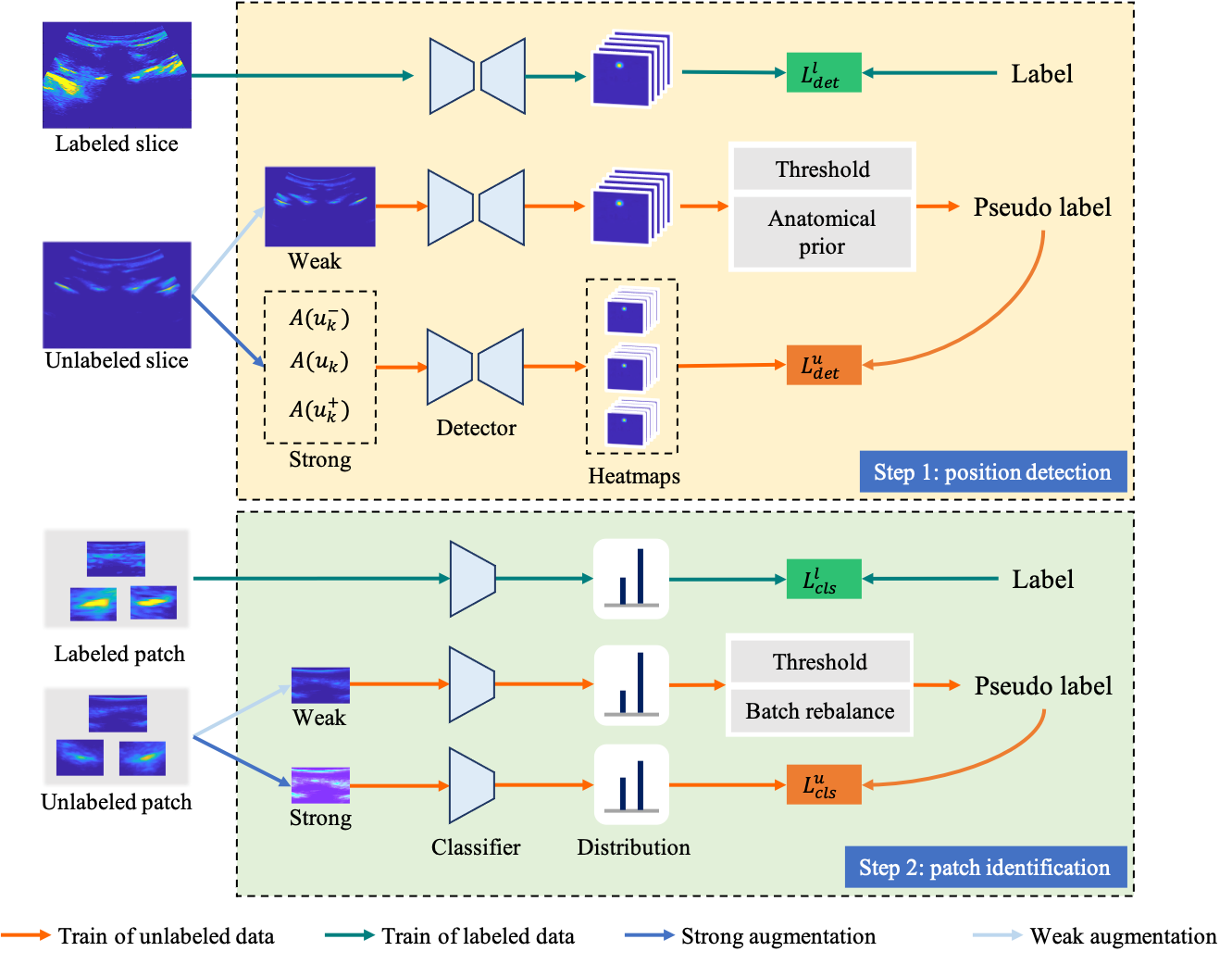}
        \captionsetup{font={small}}
    \caption{Overview of the two-step VertMatch framework for vertebral structures detection. The first step is to detect the possible positions of vertebral structures as five landmarks, and then the patches are cropped based on the detection results; the second step uses a classifier to distinguish whether the patches contain real vertebral structures. Both detector and classifier are trained via similar semi-supervised manner using the labeled and unlabeled data. In the first step, anatomical prior is used to select better pseudo labels, and the unlabeled slice ($u_k$) and its adjacent slices ($u_k^-,u_k^+$) are strong-augmented ($A(\cdot)$) and input together to utilize more unlabeled data based on proposed multi-slice consistency. In the second step, the batch rebalance is used to balance the number of pseudo labels in different categories in a batch.}
    \label{fig:method}
\end{figure}

The overview of VertMatch is shown in Fig. \ref{fig:method}. The first step is to detect possible positions of structures on transverse slices.  The detection network is trained following a semi-supervised manner by utilizing multi-slice consistency, and anatomical prior is used to select more credible pseudo labels during training. Afterwards, the patches are cropped based on the detected positions. The second step is to identify whether these patches contains real vertebral structures. In this step, the SSL algorithm is deployed with batch rebalance to solve the imbalanced data problem. 

\subsection{Position detection}
\subsubsection{Landmark detection}
The possible position of SP and laminae are detected as five landmarks. The detector uses the SHN (\cite{newellStackedHourglassNetworks2016}) or HRNet (\cite{wangDeepHighResolutionRepresentation2021}) as backbone network to extract feature map, and uses $3\times3$ convolution layers to process feature map and output predicted heatmaps. 

Given an input slice $I\in\mathbbm{R}^{W\times H\times 1}$ with width $W$ and height $H$, its annotated five landmarks are encoded to a set of heatmaps $h\in[0,1]^{\frac{W}{d}\times \frac{H}{d}\times 5}$ by using a Gaussian kernel centering at landmarks, where $d$ is downsampling ratio and empirically set as 4. The detector $f_\theta$ outputs predicted heatmaps $\hat{h}\in[0,1]^{\frac{W}{d}\times \frac{H}{d}\times 5}$ and the mean squared error (MSE) loss $L^l_{det}$ is calculated between two sets of heatmaps. Note that the coordinates of predicted landmarks need to be decoded from the output heatmaps and restored to original resolution. Let $\textit{X}=\{(x_k,h_k): k\in(1,...,N)\}$ be $N$ labeled slices with heatmap annotation $h_k$ in a batch, and the supervised MSE loss can be written as:
\begin{equation}
\label{det_l}
    L^l_{det}=\frac{1}{N}\sum_{k=1}^N \big\|h_k-\hat{h_k}\big\|^2
\end{equation}
where $\hat{h_k}=f_\theta(x_k)$ is the predicted heatmaps when inputting $x_k$.

\subsubsection{SSL-based position detection}
To avoid large annotations, we proposed a novel SSL method to train the detector as shown in Fig. \ref{fig:method} Step 1. Following basic SSL training procedure, the detector is trained by using labeled (green arrow) and unlabeled data (orange arrow) simultaneously in each batch. For labeled slice, the supervised MSE is calculated directly as illustrated in Eq. \ref{det_l}. For unlabeled slice, the pseudo labels are generated from weakly-augmented slice, and then the detector is trained by forcing to output similar predictions when being fed strongly-augmented slice. Especially, we use anatomical prior to improve quality of pseudo labels, and propose multi-slice consistency to utilize more unlabeled data.
\begin{figure}[t]
    \centering
    \includegraphics[scale=0.7]{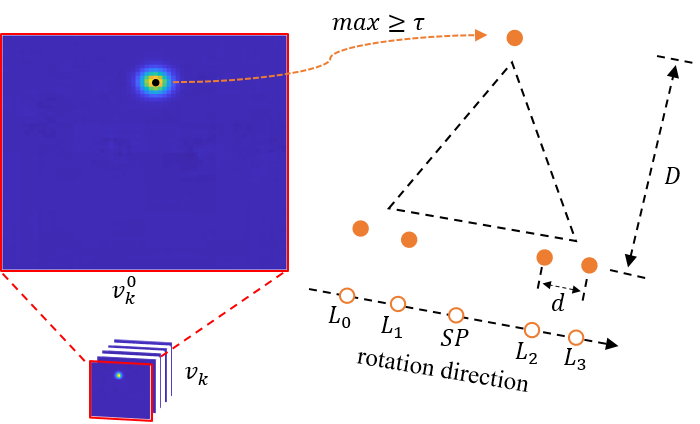}
    \captionsetup{font={small}}
    \caption{Generating standard-compliant pseudo label. Firstly, coordinates of all five landmarks are decoded from heatmaps $v_k$. The left image shows decode procedure of SP heatmap $v_k^0$ by finding the maximal activation above threshold $\tau$, and the black dot is the maximal activation. Secondly, as shown in the right image, all decoded landmarks (orange dots) should obey anatomical prior: 1) forming acute triangle; 2) following specific order along rotation direction; 3) $D$ and $d$ are within a reasonable range.}
    \label{fig:pseudo}
\end{figure}

We uses both confidence threshold and anatomical prior to generate better pseudo labels as illustrated in Fig. \ref{fig:pseudo}. Given a unlabeled slice $u_k$, the weak augmentation $\alpha(\cdot)$ is applied and then the detector outputs heatmaps $v_k=f_\theta(\alpha(u_k))$. One landmark is decoded from one heatmap $v_k^i$ by finding the maximal activation above threshold $\tau$, moreover the unlabeled slice $u_k$ will be abandoned directly if there are a maximal activation less than threshold $\tau$. Afterwards, five decoded landmarks will be accepted as pseudo label if they obey anatomical prior. According to prior acknowledge, the basic shape of vertebra and the relative position between SP and laminae are consistent, and sizes of various vertebrae are usually within reasonable range for different subjects. Therefore, anatomical prior can be summarised into three aspects: 1) five landmarks should form an isosceles acute triangle approximately; 2) five landmarks should follow the order $\{L_0,\ L_1,\ SP,\ L_2,\ L3\}$ along rotation direction; 3) the distance $D$ between SP and laminae and length $d$ of two laminae should be within reasonable range. Anatomical prior can screen pseudo labels produced by confidence threshold, and provide high quality pseudo labels for unlabeled data.

Consistency regularization assumes that the predictions $f_\theta(\alpha(u_k))$ and $f_\theta(A(u_k))$ should be similar when fed perturbed versions of one slice, where $A(\cdot)$ is the strong augmentation. We further proposed multi-slice consistency, which declares that the possible positions of vertebral structures of adjacent slices do not mutate, and the detector should output similar predictions when being fed perturbed versions of the adjacent slices. Let $u_k^-$ and $u_k^+$ be adjacent slices of $u_k$, the detector will output three sets of heatmaps $\hat{v_{k,1}}=f_{\theta}(A(u_k^-)),\hat{v_{k,2}}=f_{\theta}(A(u_k)),\hat{v_{k,3}}=f_{\theta}(A(u_k^+))$, and each set of heatmaps should be similar to pseudo label $v_k=f_\theta(\alpha(u_k))$. Multi-slice consistency can provide more unlabeled slices for training and obtain better detector. 

Let $\textit{U}=\{u_k: k\in(1,...,\mu N)\}$ be $\mu N$ unlabeled slices in a batch, where $\mu$ is a hyperparameter that determines the
relative sizes of $X$ and $U$. The unsupervised loss and final semi-supervised detection loss can be written as:
\begin{equation}
\label{det_u}
    L^u_{det} = \frac{1}{\mu N}\sum_{k=1}^{\mu N}\psi(v_k)\sum_{j=1}^3\sum_{i=1}^5\mathbbm{1}\Big\{max(v_k^i)\geq\tau\Big\}\ \big\|v_k^i-\hat{v_{k,j}^i}\big\|^2
\end{equation}
\begin{equation}
\label{det}
    L_{det} = L^l_{det}+\lambda_u L^u_{det}
\end{equation}
where the $\psi(\cdot)=1$ if predicted five landmarks obey anatomical prior, else $\psi(\cdot)=0$. The $\lambda_u$ is a hyper-parameter used to weight the unlabeled loss.

\subsection{Patch generation}
\begin{figure}[t]
    \centering
    \includegraphics[width=0.5\linewidth]{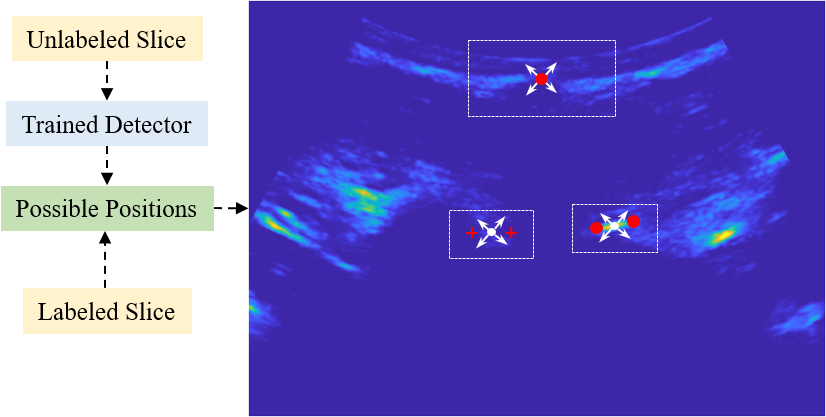}
    \captionsetup{font={small}}
    \caption{Cropping patches using random center on labeled and unlabeled slices. Three patches are cropped on one slice centering at SP position and two middle points (white dots) of two lamina landmarks. The white arrows mean that the centers of patches will move in small range in every epoch.}
    \label{fig:patch}
\end{figure}

The local patches are obtained based on possible positions as illustrated in Fig. \ref{fig:patch}. The possible positions can be obtain directly from annotations for labeled data, and need to be decoded from outputted heatmaps of detector for unlabeled data. There is no threshold for this decode procedure, and the maximal activation on each heatmap is regarded as the predicted landmark directly.

Normally, the middle point of two lamina landmarks are calculated, and three patches are finally obtained on one transverse slice by cropping rectangle areas centering at the SP position and two middle points. According to experiment, the SP patch is set to 110$\times$140 pixels and the laminae patch is set to 80$\times$100 pixels. Compared to fixed centers of patches, we move the centers randomly within small range during each training. The unfixed centers can be regarded as data augmentation, and could eliminate the influence of imprecise detection.

\subsection{Patch identification}
After obtaining patches, the classifier is used to identify whether the patches contain real vertebra structures as shown in Fig. \ref{fig:method} Step 2. We uses ResNet18 as the backbone, and train the classifier in semi-supervised manner by utilizing $N$ labeled patches and $\mu N$ unlabeled patches in each batch. For labeled patches (green arrow), the standard cross-entropy loss is applied directly:
\begin{equation}
    L_{cls}^l = \frac{1}{N}\sum_{k=1}^N H(y_k,p_k)
    \label{det}
\end{equation}
where $y_k$ is category label and $p_k$ is the predicted probability distributions, $H$ is cross-entropy loss.

For an unlabeled patch $I_k$ (orange arrow), the weak augmentation $\alpha(\cdot)$ and the strong augmentation $A(\cdot)$ are applied and then the classifier $g_\theta$ outputs two probability distributions $q_k=g_\theta(\alpha(I_k))$ and $\hat{q_k}=g_\theta(A(I_k))$. Normally, the pseudo label is produced using a high confidence threshold $\tau$, and the cross-entropy loss is calculated between pseudo label and $\hat{q_k}$ to train classifier according to consistency regularization.

However, the categories in unlabeled patches are imbalanced, since the slices with vertebral structures are minority in ultrasound scan, especially the SP only appears on small amount of slices. The imbalanced unlabeled patches will cause the bias toward majority classes when generating pseudo labels. Therefore, we rebalance the categories of pseudo labels in a batch, which means the number of majority is forced to be equal to the number of minority category. Let $M$ be a mask array with $\mu N$ elements, and each element $m_k=\mathbbm{1}\Big\{max(\hat{q_k})\geq\tau)\Big\}$ is initialized based on confidence threshold. The rebalance operation can be presented as:
\begin{equation}
    M = {Reb}(M,n)
    \label{eqa:rebalance}
\end{equation}
where $n$ is the number of minority category. $Reb(\cdot)$ is the rebalance operation putting on mask array $M$, to randomly keep $n$ elements belonging to majority category. The unsupervised loss and classification loss can be written as:
\begin{equation}
    L_{cls}^u = \frac{1}{\mu N}  \sum_{k=1}^{\mu N} m_k\ H(z_k,\hat{q_k})
    \label{eqa:cls_u}
\end{equation}
\begin{equation}
    L_{cls} = L^l_{cls}+\lambda_u L^u_{cls}
    \label{eaq:cls}
\end{equation}
where $z_k=\arg\max(q_k)$ is category prediction.

\section{Experiments}
\subsection{Subjects and datasets}
The ultrasound spine data are collected from 90 adolescents (17 males and 73 females; mean age:14.2±2.0; body mass index: 14.3$\sim$22.7). They fulfill the following inclusion criteria: 1) no constraints surgical treatment; 2) receive a posterior-anterior standing radiograph and US scan within an hour; 3) with the major curve of 10° to 45°. Before being enrolled into the study, all participating subjects signs a written consent, and ethics approval is granted from the local health research ethics board. The radiographic curve information including SPAs and Cobb angles are exported from the medical record to be used for the following data analysis. 

One Standing ultrasound spine scan contains approximate 900-2300 transverse slices acquired along the spine curve, and is obtained using the SonixONE system equipped with 128-element C5-2/60 array transducer and SonixGPS (Analogic Ultrasound - BK Medical, Peabody, Massachusetts, US). The center frequency of transducer probe is 2.5 MHz, and the imaging depth is set as 6-8cm. 

Fifty ultrasound scans are used to build the training and testing datasets, and another forty ultrasound scans are applied for clinical application. The detailed information of our datasets is illustrated in Table.\ref{tab:data}. For the training labeled data, we conditionally selected slices in each ultrasound scan, and try to maintain the category balance between slices with and without vertebral structures. For testing data, it is not necessary to annotate all slices in subject scans. In order to cover more cases, we sample every ten slices in 10 subjects scans, and randomly select 20 slices in another 10 subjects scan. 
\begin{table}[htbp]
\small
\begin{center}
    \caption{Information of datasets built from fifty ultrasound scans.}
    \label{tab:data}
    \begin{tabular}{c|c|cc}
        \toprule
        \multicolumn{2}{c|}{Dataset} & Subject Number & Slice Number \\
        \midrule
        \multirow{2}*{train}&labeled &10 & 1000 \\
        \cmidrule{2-4}
        &unlebeled & 20 & 26510\\
        \midrule
        \multicolumn{2}{c|}{test} & 20 & 1675  \\
        \bottomrule
    \end{tabular}
\end{center}
\end{table}

\subsection{Evaluation metrics}
We use the Percentage of Correct Keypoints (PCK) to measure the performance of position detection in first step, which reports the percentage of detected landmarks that are within the region of the ground truth, and could expressed as:
\begin{equation}
    PCK=\frac{\sum_n\sum_5\delta(d\leq T)}{n}
    \label{eq3}
\end{equation}
where $n$ is number of landmarks, and $d$ is the distance between the predicted landmark and ground-truth. The $\delta(\cdot)=1$ when distance less than threshold $T$, which is set to 15 pixels and 30 pixels for real and fake vertebral structures respectively. The landmarks of fake vertebral structures are hard to match the ground-truth perfectly due to inapparent features, and the large threshold is acceptable since they will be excluded by classifier. 

We use the accuracy, precision, recall and F1-score to evaluate the entire method, which are widely-used metrics in classification task. Note that these metrics are redefined in our study, and the true positive (TP) and true negative (TN) are defined as:
\begin{equation}
\begin{split}
     &TP: \mathbbm{1}\{y=1\}\ and \ \delta(d\leq T) \\ 
     &TN: \mathbbm{1}\{y=0\}
    \label{eq_tn}
\end{split}
\end{equation}
where the newly defined TP requests both position prediected by detector and category predicted by classifier are correct. 

\subsection{Implmentation details}
All experiments are implemented using Pytorch on two 24 GB NVIDIA RTX 3090 GPUs. Both the detector and classifier are trained from scratch with 500 epochs. The detector uses the Adam optimizer, and initial learning rate is 1e-3 with decay schedule of 0.1 in 60, 150 epoch. We set $N=8$, $\mu=3$, $\tau=0.8$, $\lambda_u=1$ for detector SSL. The classifier uses the SGD optimizer, and the learning rate is initialized to 3e-2 and decays by 0.1 after 200 epochs. The momentum is 0.9 and the weight decay is 5e-4. We set $N=32$, $\mu=15$, $\tau=0.97$, $\lambda_u=1$ for classifier SSL.

The weak and strong data augmentations are important for the semi-supervised learning, and have been widely used in SSL method (\cite{sohnFixMatchSimplifyingSemiSupervised2020}\cite{berthelotMixMatchHolisticApproach2019a}). In our study, the weak augmentations are random Gaussian noise augmentation for detector, and random horizontal flip augmentation for classifier. We use strong augmentations based on RandAugment \cite{sohnFixMatchSimplifyingSemiSupervised2020}, but adjust some parameters and delete color augmentation to fit the ultrasound data. Besides, we abandon all augmentations about spatial transform when training detector, since it is difficult to perform the same spatial transforms for the landmarks and the results showed no influence without them.

\subsection{Quantitative and qualitative analysis}
\textbf{Quantitative Comparison with Alternatives}
To evaluate our proposed method, considering there are no existing work for this task, we compare it with some one-step and two-step methods as shown in Table. \ref{tab_compare}. The one-step methods use two well-known detection networks SHN (\cite{newellStackedHourglassNetworks2016}) and HRNet (\cite{wangDeepHighResolutionRepresentation2021}) to directly predict real vertebral structures. The other two-step methods combine SHN/HRNet and ResNet18, and consist of possible position detection and patch identification following supervised manner. VertMatch also uses the SHN and HRNet as backbone network in detector, and the performances of two backbones are compared. All methods are trained using labeled data, and VertMatch utilizes unlabeled data additionally. 

Table.\ref{tab_compare} lists the quantitative results of different methods. The one-step supervised methods show unsatisfied results especially in SP detection, and the two-step supervised methods attains significant improvements in accuracy and f1-score. Two-step methods include global position detection and local patch identification, and these improvements demonstrate that the decomposition of entire task can truly reduce difficulty. VertMatch using SHN achieves the best performance in accuracy and f1-score. Compared to SHN+ResNet18, the accuracy and f1-score increase by 2.5\%, 3.4\% of SP detection and 1.7\%, 2.5\% of laminae detection respectively, which illustrates that VertMatch can utilize the unlabeled data efficiently and reach better results. Since the SHN outperforms the HRNet in most results, we use the SHN as the backbone in the following experiments. Besides, all methods achieve a better performance for lamina detection, since the laminae have more notable characters compared to SP.
\begin{table*}[ht]
\small
\begin{center}
    \caption{Quantitative comparison between different methods. VertMatch is semi-supervised and uses 1000 labeled and 26510 unlabeled slices, and other methods only use 1000 labeled slices.}
    \label{tab_compare}
    \begin{tabular}{cccccccccc}
        \toprule
        \multicolumn{2}{c}{\multirow{2}{*}{Method}} & \multicolumn{4}{c}{Spinous Process} & \multicolumn{4}{c}{Laminae}\\
        \cmidrule(r){3-6} \cmidrule(r){7-10}
                                    &                & Accuracy & Precision & Recall & F1-Score &Accuracy & Precision & Recall & F1-Score \\
        \midrule
        \multirow{2}{*}{One-step}   &HRNet           &71.6   &47.5   &64.6   &54.7   &81.3   &65.2   &92.6   &76.5\\
                                    &SHN             &63.4   &41.7   &98.9   &58.6   &82.1   &65.6   &97.6   &78.4\\
        \midrule
        \multirow{4}{*}{Two-step}   &HRNet+ResNet18  &85.2   &68.8   &84.4   &75.8   &89.9   &83.9   &86.3   &84.5\\
                                    &SHN+ResNet18    &86.9   &72.7   &85.8   &78.7   &89.6   &83.0   &86.8   &84.8\\
                                    &VertMatch(HRNet)      &87.9   &74.2   &85.7   &79.2   &89.1   &79.1   &90.8   &85.1\\
                                    &VertMatch(SHN)        &89.4   &77.7   &87.0   &82.1   &91.3   &83.6   &92.2   &87.3\\
        \bottomrule
    \end{tabular}
\end{center}
\end{table*}

\textbf{Quantitative Analysis of Labeled Data Volume}
To explore the impact of labeled data volume on VertMatch, the evaluation is performed with different number of labeled data and the number of used data decreases with the number of subjects. As illustrated in Table.\ref{tab_semi}, the performance of the VertMatch is promoted when more labeled data are available, and there is no distinctive upgrade between results from 700 and 1000 slices, which means that the labeled data play a limited role unless the number of data is massively scaled up. When using 700 labeled slices, VertMatch still outperforms all supervised two-step methods using 1000 labeled slices, and the SP detection achieves 2.8\% and 3.5\% improvement in accuracy and f1-score respectively compared to SHN+ResNet18. It demonstrates that VertMatch is reliable for vertebral structures detection although without a large amount of labeled data, and our semi-supervised technique can learn the acknowledge from unlabeled data efficiently. 
\begin{table*}[ht]
\small
\begin{center}
    \caption{Quantitative analysis of different number of labeled data when using VertMatch.}
    \label{tab_semi}
    \begin{tabular}{c c c c c c c c c}
        \toprule
        \multirow{2}{*}{Labeled / Unlabeled} & \multicolumn{4}{c}{Spinous Process} & \multicolumn{4}{c}{Laminae}\\
        \cmidrule(r){2-5} \cmidrule(r){6-9}
         & Accuracy & Precision & Recall & F1-Score &Accuracy & Precision & Recall & F1-Score \\
        \midrule
        100(1) / 26510  &77.6   &56.5   &67.4   &61.5   &86.0   &74.5   &87.6   &80.5\\
        300(3) / 26510  &81.9   &62.9   &76.1   &68.9   &85.7   &74.3   &86.2   &79.8\\
        500(5) / 26510  &83.5   &65.8   &82.9   &73.2   &86.9   &76.1   &89.0   &82.0\\
        700(7) / 26510  &89.7   &79.6   &84.8   &82.1   &89.5   &81.3   &89.2   &85.1\\
        1000(10) / 26510  &89.4   &77.7   &87.0   &82.1   &91.3   &83.6   &92.2   &87.3\\
        \bottomrule
        \multicolumn{9}{@{}l}{$^*$300(3) indicates 300 slice are selected from 3 ultrasound scans}
    \end{tabular}
\end{center}
\end{table*}

\begin{figure*}[ht]
    \centering
    \includegraphics[scale=0.64]{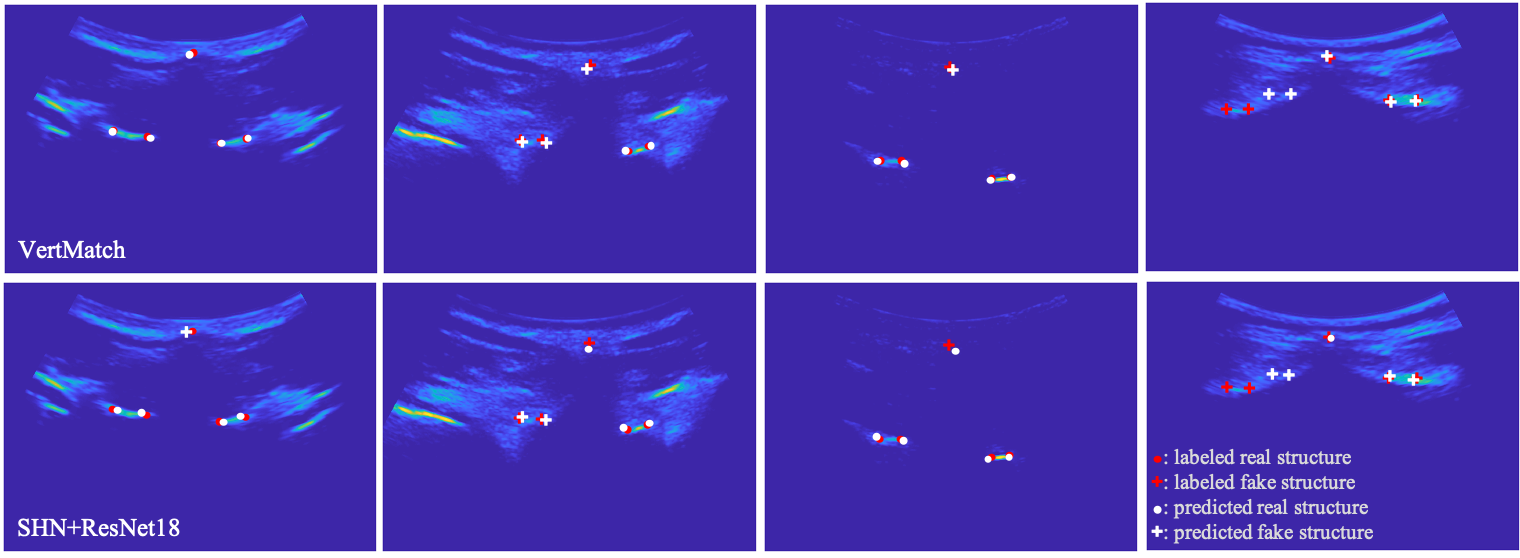}
    \captionsetup{font={small}}
    \caption{Visual comparison on challenging transverse slices. AS shown in the $1^{st}$-$3^{rd}$ columns (left to right), the SHN+ResNet18 method can detect possible positions accurately, but cannot distinguish real structures (white cross $vs$ red dot). On the fourth column, the classifier can exclude the fake structures although the predicted landmarks (white cross) are far from labeled landmarks (red cross).}
    \label{fig:vis_trans}
\end{figure*}

\textbf{Qualitative Comparison}
We compare the visual results of some challenging transverse slices using SHN+ResNet18 and VertMatch with SHN, respectively. As shown in Fig. \ref{fig:vis_trans}, four transverse slices from different subjects are detected, and it is observed that the VertMatch is better at detecting the real SP and handling the complexity of ultrasound data. Both two methods can detect the possible position of SP and laminae accurately, but the supervised method cannot distinguish the real structures well due to limited labeled data. On the fourth slice, the predicted landmarks of left lamina are far from annotated possible positions but can be excluded by the classifier.

\begin{figure*}[ht]
    \centering
    \begin{minipage}[t]{0.33\linewidth}
        \centering
        \includegraphics[scale=0.54]{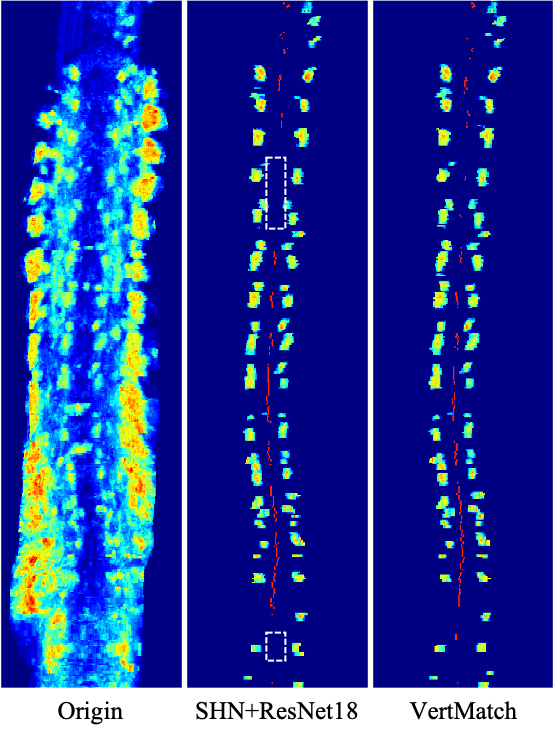}
    \end{minipage}
    \begin{minipage}[t]{0.33\linewidth}
        \centering
        \includegraphics[scale=0.54]{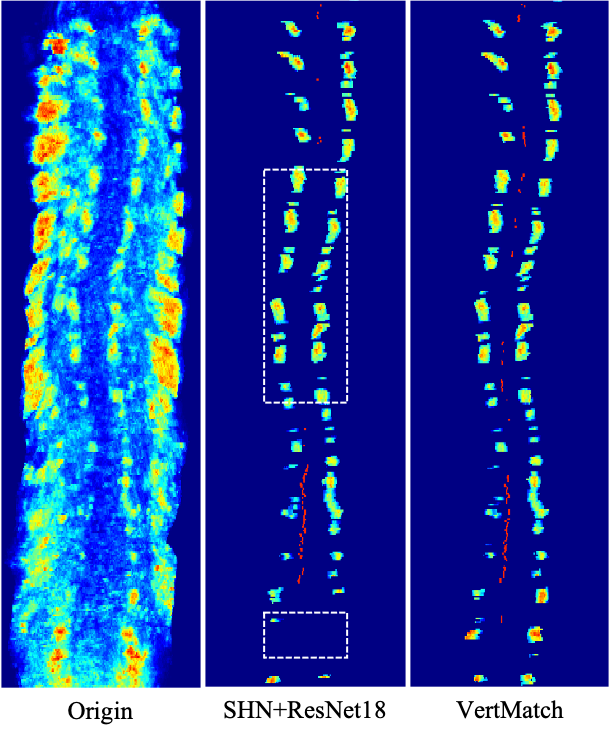}
    \end{minipage}
    \begin{minipage}[t]{0.33\linewidth}
        \centering
        \includegraphics[scale=0.54]{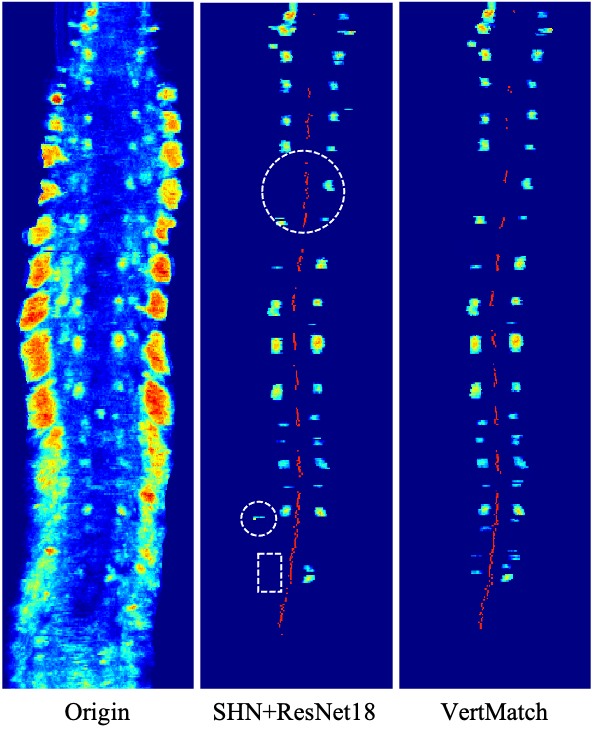}
    \end{minipage}
    \captionsetup{font={small}}
    \caption{Visual comparison on coronal projection images of ultrasound scans. The red dots are detected SPs, which are visualized on coronal image. The dotted boxes reveal missing detection of SP or lamina, the dotted circles reveal wrong or redundant detection.}
    \label{fig:vis_coronal}
\end{figure*}

Besides on transverse slices, we also compared these two methods on whole ultrasound scans. For entire scan, the SP and laminae are detected firstly on all transverse slices using two detection methods. Based on the predicted coordinates, a rectangular region around the detected laminae are segmented to remove ribs and soft tissue layer; the coordinates of the detected SP and its 8-neighbor pixels are assigned to 255. Subsequently, the Voxel-Based Nearest Neighbor (VNN) and Fast Dot Projection (FDP) method (\cite{chenImprovement3DUltrasound2021}) are used to reconstruct the US scan to obtain the projected coronal image. As shown in Fig. \ref{fig:vis_coronal}, the laminae and SPs are clearly separated in pairs, and the SPs are visualized on coronal plane to provide more information compared to original image. At the areas marked by the dotted boxes, VertMatch represents more detected structures which can expose the clear posterior anatomy of spine; at the areas marked by the dotted circles, VertMatch achieves better classification results and the pairs of SP and laminae are more obvious. Both two methods perform worse in lumbar region, since ultrasound signals are largely attenuated by thick and multi-layered soft tissue, and SP and laminae are very difficult to detect on transverse slice. 

\subsection{Ablation study}
We conduct the ablation experiments to validate the effectiveness of the proposed techniques of semi-supervised learning, and all experiments are implemented using 700 labeled data and all unlabeled data.

\textbf{Anatomical Prior} The influence of anatomical constraint on position detection is explored by calculating the PCK as shown in Table. \ref{tab:pts}. Anatomical prior is used to select better pseudo labels, and detector benefits from high-quality pseudo labels in SSL training and achieves 2.6\% improvement of PCK.

\textbf{Multi-Slice Consistency} Based on assumption that possible positions of structures are consistent on adjacent slices, multi-slice consistency can utilize more unlabeled data by inputting three consecutive slices. As shown in Table. \ref{tab:pts}, the position detection PCK raises 2.6\% using multi-slice consistency, since more unlabeled data can train a detector with better generalization. 

More importantly, the detector using both anatomical prior and multi-slice consistency achieves remarkable improvement of 5.4\% in PCK, indicating complementary advantage of two techniques. 
\begin{table}[ht]
\small
\begin{center}
    \caption{Influence of multi-slice consistency and anatomical prior on position detection.}
    \label{tab:pts}
    \begin{tabular}{p{2cm}<{\centering} p{2cm}<{\centering}|p{1cm}<{\centering}}
        \toprule
        Anatomical prior & Multi-Slice Consistency & PCK(\%) \\
        \midrule
        \ding{55} &\ding{55} & 79.5\\
        \checkmark & \ding{55} & 82.1\\
        \ding{55} & \checkmark & 82.1\\
        \checkmark & \checkmark & 84.9\\
        \bottomrule
    \end{tabular}
\end{center}
\end{table}

\textbf{Batch Rebalance \& Random Center} 
The random center and batch rebalance are used to crop and identify patch, and their influence is investigated as illustrated in Table.\ref{tab:cls}. If we use random center to obtain patches, the accuracy and f1-score of laminae detection increase by 3.8\% and 3.4\%, but the SP detection results only improve slightly. If the batch rebalance is contained, both SP and laminae detection are improved and achieve 3.8\% and 2.0\% increase in f1-score, which proves that more true position are predicted and the batch rebalance helps to solve the imbalanced problem. Finally, if we add both two techniques into basic method, the final results achieve the highest accuracy and f1-score among all the four methods, which demonstrate the effectiveness of the random center and batch rebalance.
\begin{table*}[ht]
\small
\begin{center}
    \caption{Influence of random center and batch rebalance.}
    \label{tab:cls}
    \begin{tabular}{c c c c c c c c c c}
        \toprule
        \multirow{2}{*}{\makecell[c]{Random\\Center}} & \multirow{2}{*}{\makecell[c]{Batch\\Rebalance}}& \multicolumn{4}{c}{Spinous Process} & \multicolumn{4}{c}{laminae}\\
        \cmidrule(r){3-6} \cmidrule(r){7-10}
        & & Accuracy & Precision & Recall & F1-Score &Accuracy & Precision & Recall & F1-Score \\
        \midrule
        \ding{55}&\ding{55}     &85.7   &77.5   &69.4   &73.2   &84.5   &69.4   &95.5   &80.4\\
        \checkmark&\ding{55}    &86.4   &80.9   &67.5   &73.6   &88.3   &78.0   &90.1   &83.8\\
        \ding{55}&\checkmark    &86.8   &75.1   &79.0   &77.0   &86.8   &74.1   &92.9   &82.4\\
        \checkmark&\checkmark   &89.7   &79.6   &84.8   &82.1   &89.5   &81.3   &89.2   &85.1\\
        \bottomrule
    \end{tabular}
\end{center}
\end{table*}

\subsection{Clinical application}
\begin{figure*}[htb]
    \centering
    \begin{minipage}[t]{0.126\linewidth}
        \centering
        \includegraphics[width=\linewidth]{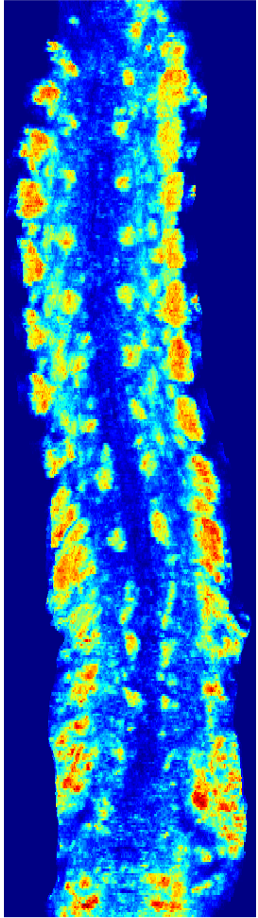}
        \subcaption{}
        \label{fig:spa_a}
    \end{minipage}     
    \begin{minipage}[t]{0.1255\linewidth}
        \centering
        \includegraphics[width=\linewidth]{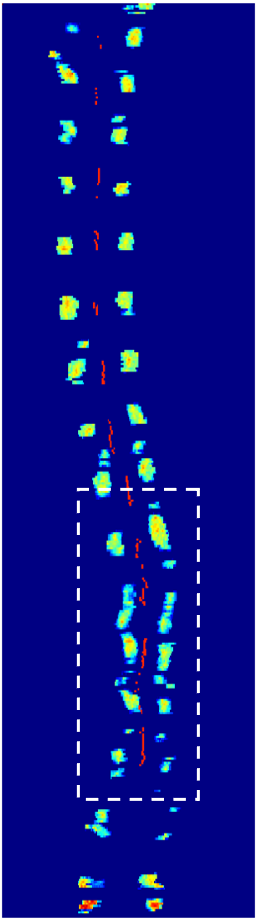}
        \subcaption{}
        \label{fig:spa_b}
    \end{minipage}
    \begin{minipage}[t]{0.126\linewidth}
        \centering
        \includegraphics[width=0.82\linewidth]{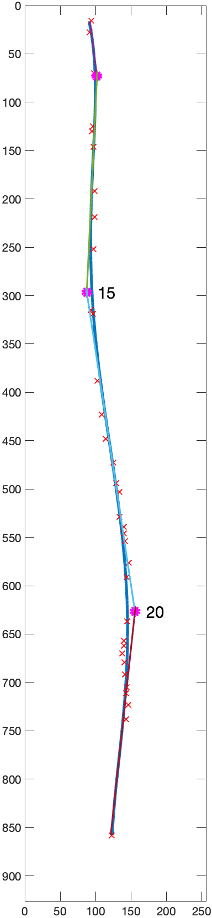}
        \subcaption{}
        \label{fig:spa_c}
    \end{minipage} 
    \begin{minipage}[t]{0.126\linewidth}
        \centering
        \includegraphics[width=0.85\linewidth]{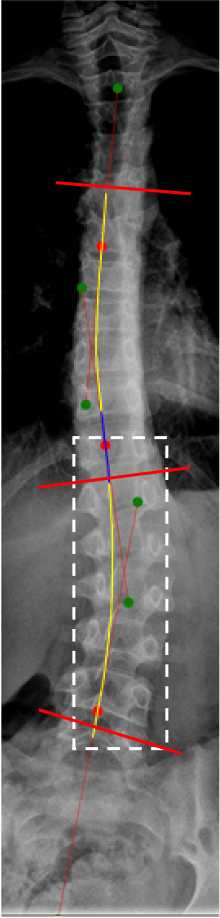}
        \subcaption{}
        \label{fig:spa_d}
    \end{minipage} 
    \begin{minipage}[t]{0.15\linewidth}
        \centering
        \includegraphics[width=\linewidth]{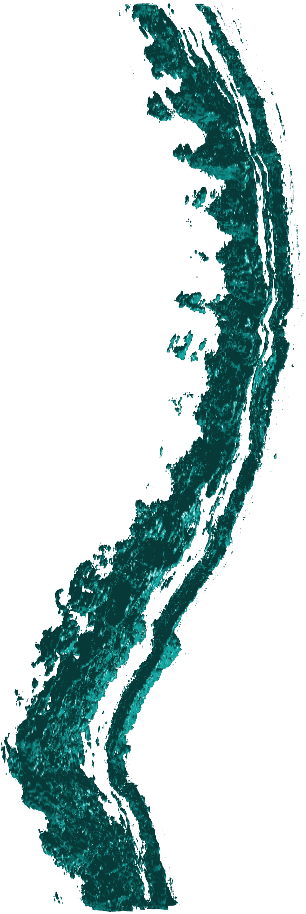}
        \subcaption{}
        \label{fig:3d_a}
    \end{minipage} 
    \begin{minipage}[t]{0.15\linewidth}
        \centering
        \includegraphics[width=\linewidth]{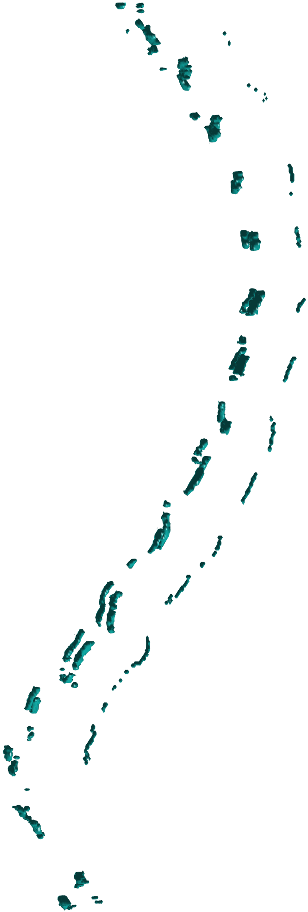}
        \subcaption{}
        \label{fig:3d_b}
    \end{minipage}
    \captionsetup{font={small}}
    \caption{The demonstration of SPA measurement and 3D rendering result: (a) original ultrasound coronal image. (b) the processed coronal ultrasound image with SP and laminae. (c) the automatic measurement on ultrasound image, and (d) the manual SPA measurement (yellow curve) on radiograph. The SPAs are 15/20° on (c) and 14/22° on (d). The Cobb angles (indicated by red lines) on (d) are 22/27°. (e)-(f) the sagittal views of 3D rendering result of original and processed ultrasound volume.}
    \label{fig:spa}
\end{figure*}
\begin{figure}[t]
    \begin{minipage}[t]{0.48\linewidth}
        \centering
        \includegraphics[width=0.6\linewidth]{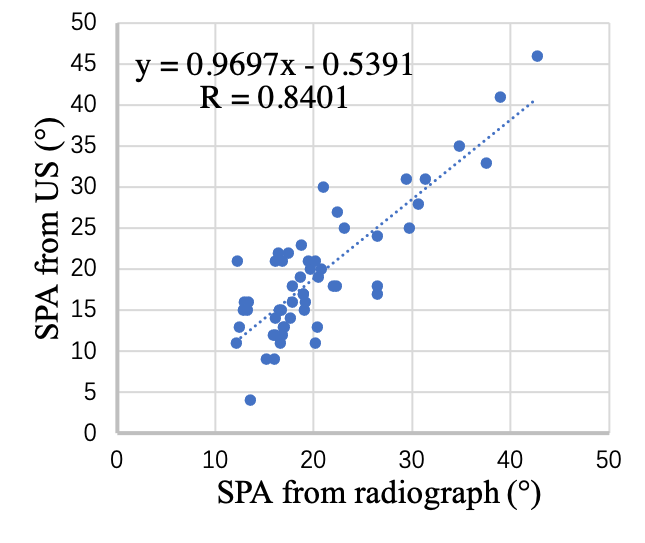}
        \subcaption{}
        \label{fig:tab_a}
    \end{minipage}     
    \begin{minipage}[t]{0.48\linewidth}
        \centering
        \includegraphics[width=0.6\linewidth]{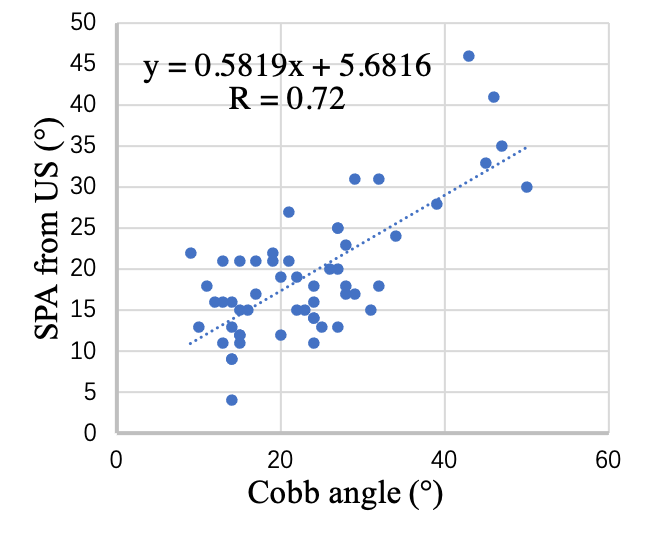}
        \subcaption{}
        \label{fig:tab_b}
    \end{minipage}
    \captionsetup{font={small}}
    \caption{Comparison results of spinal curvature measurements on forty ultrasound scans: (a) The comparison of SPAs between ultrasound and radiographic measurements of SPAs. (b) The comparison between SPAs from US images and Cobb angles.}
    \label{fig:tab}
\end{figure}
We further investigate the potential of VertMatch on AIS assessment. The Cobb angles is usually used to diagnose and monitor AIS (\cite{morrissyMeasurementCobbAngle1990}; \cite{zhouAutomaticMeasurementSpine2017}), but  it only provides the information of lateral spinal curvature. The spinous process angle (SPA) is an alternative measurement parameter which can reveal the information of both the axial vertebral rotation and lateral curvatures (\cite{herzenbergCobbAngleSpinous1990}). However, the spinous process is usually merged into soft tissue and invisible on ultrasound coronal images, hence the SPA cannot be measured on ultrasound image.

VertMatch can visualize the SP on coronal spine image by detecting SP position on transverse slices of ultrasound scan, and the SPA can be measured automatically on the coronal image (\cite{zengAutomaticDetectionMeasurement2021}; \cite{geAutomaticMeasurementSpinous2020}) as shown in Fig. \ref{fig:spa_c}.  Forty ultrasound scans outside the training and testing datasets are applied for the SPA measurement, fifty-five SP curves are obtained and compared to the SPAs measured from radiographs.

Fig. \ref{fig:spa_b} \& fig. \ref{fig:spa_c}\&\ref{fig:spa_d} illustrated an example of SPA measurement from two modalities. As shown in the white box, the SP curves reveal apparent vertebral rotation and are deviated to the same direction (from the middle line of vertebra to the left side) on both ultrasound images and radiograph. It indicated that the SP curves obtained by ultrasound qualitatively keep the similar trend as radiography. Fig. \ref{fig:tab_a} compared the SPA measurements between two modalities quantitatively. The automatic measurements show high agreement ($R=0.84$) to radiographic results, and the average measurement differences is $3.63\pm2.60^\circ$ and less than the clinical acceptance error ($5^\circ$). Fig. \ref{fig:3d_b} shows the 3D rendering result of process volume, and the SP and laminae are clearly separated for different vertebral levels compared to Fig. \ref{fig:3d_a}. All the results illustrate that the VertMatch is a reliable and promising tool for SPA measurement and AIS assessment.

Fig. \ref{fig:tab_b} illustrates the comparison between SPAs from ultrasound measurements and Cobb angles from radiographs, and the moderate correlation between two angles indicates that there is inconsistency between these two evaluation parameters of scoliotic curves. As show in Fig. \ref{fig:spa_d}, the Cobb angle is used to reveal the spinal curvature, which is apparently different from yellow SP curve. The SPA usually is smaller compared to Cobb angle especially for large curvature, because the vertebrae generally rotate to the curve convexity and thus cause the SP tips offset laterally to the direction of the curve concavity (\cite{morrisonCorrelationCobbAngle2015}).

\section{Discussion and conclusion}
The 3D Ultrasound imaging technique has become a promising diagnostic tool for AIS, since it is radiation-free, cost-effective and portable comparing to radiography. However, current 3D ultrasound volume is only applied to measure the spinal curvature by using coronal projection image, and it is necessary to detect vertebra in ultrasound volume to reveal 3D deformity of spine. In this paper, we proposed VertMatch for vertebra detection, which detect real vertebral structures on transverse slices from global to local, and utilize unlabeled data efficiently with three novel semi-supervised learning technique: multi consistency, anatomical prior and batch rebalance. VertMatch outperforms state-of-the-art methods and achieves good results using limited labeled data, furthermore it can be a promising approach for 3D assessment of scoliosis.

Multi-slice consistency inputs adjacent slices and trains the detector by forcing to output similar detection results, and has been proved to be an effective manner to utilize more unlabeled data. In this paper, we only input one slice and its two adjacent slices into detector. However, there are more than two slices which are similar to this slice, and the number of input slices is worth investigating. In the future work, we shall explore multi-slice consistency and try to input more unlabeled slices. Anatomical prior is used to generate more convinced pseudo labels. Actually, the confidence threshold is important to landmark decode, furthermore influences the performance of anatomical prior. For example, lower threshold may challenge the soft constraints from anatomical prior and ignore wrong detection in pseudo labels, higher threshold makes it difficult to decode all five landmarks and establish relationship between landmarks. The confidence threshold therefore is an essential parameter worth to be explored. 

We validate VertMatch on forty ultrasound scans for clinical application, and the SPA can be measured on processed coronal image. The results show a good correlation (0.84) between US and radiographic measurements, and automatic SPA measurement based on VertMatch can be regarded as a reliable and promising tool for the diagnosis, monitoring and screening of spine deformities. Furthermore, the vertebral structure detection in 3D ultrasound volume can be used for parametric spine modeling imaging, and it could provide clearer visualization of spine which will be comparable to CT and MRI. The individual parametric model is now generated from pose and shape parameters manually measured from 3D ultrasound volume \cite{Gao2022}. The vertebra detection technique proposed in this paper is expected to extract these parameters automatically based on the detected landmarks, and establish intuitive 3D parametric spine model in the future.

\section*{Acknowledgments}
The authors would like to extend sincere gratitude to the Glenrose Hospital for providing the subjects’ US scans in this study. And the authors also are profusely grateful to the sponsorship from Natural Science Foundation of China (NSFC), Grant No. 12074258.


\small
\bibliographystyle{unsrt}  
\bibliography{references}

\begin{thebibliography}{10}

\bibitem{weinsteinAdolescentIdiopathicScoliosis2008}
Stuart~L. Weinstein, Lori~A. Dolan, Jack~CY Cheng, Aina Danielsson, and Jose~A.
  Morcuende.
\newblock Adolescent idiopathic scoliosis.
\newblock {\em The Lancet}, 371(9623):1527--1537, May 2008.

\bibitem{asherAdolescentIdiopathicScoliosis2006}
Marc~A. Asher and Douglas~C. Burton.
\newblock Adolescent idiopathic scoliosis: Natural history and long term
  treatment effects.
\newblock {\em Scoliosis}, 1(1):2, March 2006.

\bibitem{weinsteinHealthFunctionPatients2003}
Stuart~L. Weinstein, Lori~A. Dolan, Kevin~F. Spratt, Kirk~K. Peterson, Mark~J.
  Spoonamore, and Ignacio~V. Ponseti.
\newblock Health and {{Function}} of {{Patients With Untreated Idiopathic
  Scoliosis}}: {{A}} 50-{{Year Natural History Study}}.
\newblock {\em JAMA}, 289(5):559, February 2003.

\bibitem{kimScoliosisImagingWhat2010}
Hana Kim, Hak~Sun Kim, Eun~Su Moon, Choon-Sik Yoon, Tae-Sub Chung, Ho-Taek
  Song, Jin-Suck Suh, Young~Han Lee, and Sungjun Kim.
\newblock Scoliosis {{Imaging}}: {{What Radiologists Should Know}}.
\newblock {\em RadioGraphics}, 30(7):1823--1842, November 2010.

\bibitem{lawCumulativeRadiationExposure2016}
Martin Law, Wang-Kei Ma, Damian Lau, Eva Chan, Lawrance Yip, and Wendy Lam.
\newblock Cumulative radiation exposure and associated cancer risk estimates
  for scoliosis patients: {{Impact}} of repetitive full spine radiography.
\newblock {\em European Journal of Radiology}, 85(3):625--628, March 2016.

\bibitem{huang2018}
Q.~Huang, Z.~Zeng, and X.~Li.
\newblock 2.5-{{D Extended Field-of-View Ultrasound}}.
\newblock {\em IEEE Transactions on Medical Imaging}, 37(4):851--859, April
  2018.

\bibitem{chenImprovement3DUltrasound2021}
Hong-Bo Chen, Rui Zheng, Li-Yue Qian, Feng-Yu Liu, Sheng Song, and Hong-Ye
  Zeng.
\newblock Improvement of 3-{{D Ultrasound Spine Imaging Technique Using Fast
  Reconstruction Algorithm}}.
\newblock {\em IEEE Transactions on Ultrasonics, Ferroelectrics, and Frequency
  Control}, 68(10):3104--3113, October 2021.

\bibitem{chenReliabilityAssessingCoronal2013}
Wei Chen, Edmond H.~M. Lou, Phoebe~Q. Zhang, Lawrence~H. Le, and Doug Hill.
\newblock Reliability of assessing the coronal curvature of children with
  scoliosis by using ultrasound images.
\newblock {\em J Child Orthop}, 7(6):521--529, December 2013.

\bibitem{cheungUltrasoundVolumeProjection2015}
Chung-Wai~James Cheung, Guang-Quan Zhou, Siu-Yin Law, Tak-Man Mak, Ka-Lee Lai,
  and Yong-Ping Zheng.
\newblock Ultrasound {{Volume Projection Imaging}} for {{Assessment}} of
  {{Scoliosis}}.
\newblock {\em IEEE Transactions on Medical Imaging}, 34(8):1760--1768, August
  2015.

\bibitem{zhouAutomaticMeasurementSpine2017}
Guang-Quan Zhou, Wei-Wei Jiang, Ka-Lee Lai, and Yong-Ping Zheng.
\newblock Automatic {{Measurement}} of {{Spine Curvature}} on 3-{{D Ultrasound
  Volume Projection Image With Phase Features}}.
\newblock {\em IEEE Transactions on Medical Imaging}, 36(6):1250--1262, June
  2017.

\bibitem{zhengIntraInterraterReliability2015}
Rui Zheng, Amanda~C.Y. Chan, Wei Chen, Douglas~L. Hill, Lawrence~H. Le, Douglas
  Hedden, Marc Moreau, James Mahood, Sarah Southon, and Edmond Lou.
\newblock Intra- and {{Inter-rater Reliability}} of {{Coronal Curvature
  Measurement}} for {{Adolescent Idiopathic Scoliosis Using Ultrasonic Imaging
  Method}}\textemdash{{A Pilot Study}}.
\newblock {\em Spine Deformity}, 3(2):151--158, March 2015.

\bibitem{zhengFactorsInfluencingSpinal2018}
Rui Zheng, Doug Hill, Douglas Hedden, James Mahood, Marc Moreau, Sarah Southon,
  and Edmond Lou.
\newblock Factors influencing spinal curvature measurements on ultrasound
  images for children with adolescent idiopathic scoliosis ({{AIS}}).
\newblock {\em PLOS ONE}, 13(6):e0198792, June 2018.

\bibitem{ungiAutomaticSpineUltrasound2020}
Tamas Ungi, Hastings Greer, Kyle~R. Sunderland, Victoria Wu, Zachary M.~C.
  Baum, Christopher Schlenger, Matthew Oetgen, Kevin Cleary, Stephen~R.
  Aylward, and Gabor Fichtinger.
\newblock Automatic {{Spine Ultrasound Segmentation}} for {{Scoliosis
  Visualization}} and {{Measurement}}.
\newblock {\em IEEE Trans. Biomed. Eng.}, 67(11):3234--3241, November 2020.

\bibitem{kamaliRealtimeTransverseProcess2018}
Shahrokh Kamali, Bryan Travers, Csaba Pinter, Andras Lasso, Tamas Ungi, Ben
  Church, Zachary M.~C. Baum, and Gabor Fichtinger.
\newblock Real-time transverse process detection in ultrasound.
\newblock In Robert~J. Webster and Baowei Fei, editors, {\em Medical
  {{Imaging}} 2018: {{Image-Guided Procedures}}, {{Robotic Interventions}}, and
  {{Modeling}}}, page~25, {Houston, United States}, March 2018. {SPIE}.

\bibitem{tangCNNbasedMethodReconstruct2021}
Songyuan Tang, Xu~Yang, Peer Shajudeen, Candice Sears, Francesca Taraballi,
  Bradley Weiner, Ennio Tasciotti, Devon Dollahon, Hangue Park, and Raffaella
  Righetti.
\newblock A {{CNN-based}} method to reconstruct 3-{{D}} spine surfaces from
  {{US}} images in vivo.
\newblock {\em Medical Image Analysis}, 74:102221, December 2021.

\bibitem{taoSpineTransformersVertebraLabeling2021}
Rong Tao, Wenyong Liu, and Guoyan Zheng.
\newblock Spine-{{Transformers}}: {{Vertebra Labeling}} and {{Segmentation}} in
  {{Arbitrary Field-of-View Spine CTs}} via {{3D Transformers}}.
\newblock {\em Medical Image Analysis}, page 102258, October 2021.

\bibitem{zhanRobustMRSpine2012}
Yiqiang Zhan, Dewan Maneesh, Martin Harder, and Xiang~Sean Zhou.
\newblock Robust {{MR Spine Detection Using Hierarchical Learning}} and {{Local
  Articulated Model}}.
\newblock In Nicholas Ayache, Herv{\'e} Delingette, Polina Golland, and Kensaku
  Mori, editors, {\em Medical {{Image Computing}} and {{Computer-Assisted
  Intervention}} \textendash{} {{MICCAI}} 2012}, Lecture {{Notes}} in
  {{Computer Science}}, pages 141--148, {Berlin, Heidelberg}, 2012. {Springer}.

\bibitem{wangAutomaticVertebraLocalization2021}
Fakai Wang, Kang Zheng, Le~Lu, Jing Xiao, Min Wu, and Shun Miao.
\newblock Automatic {{Vertebra Localization}} and {{Identification}} in {{CT}}
  by {{Spine Rectification}} and {{Anatomically-constrained Optimization}}.
\newblock In {\em 2021 {{IEEE}}/{{CVF Conference}} on {{Computer Vision}} and
  {{Pattern Recognition}} ({{CVPR}})}, pages 5276--5284, {Nashville, TN, USA},
  June 2021. {IEEE}.

\bibitem{xieSelfTrainingNoisyStudent2020}
Qizhe Xie, Minh-Thang Luong, Eduard Hovy, and Quoc~V. Le.
\newblock Self-{{Training With Noisy Student Improves ImageNet
  Classification}}.
\newblock In {\em Proceedings of the {{IEEE}}/{{CVF Conference}} on {{Computer
  Vision}} and {{Pattern Recognition}}}, pages 10687--10698, 2020.

\bibitem{sohnFixMatchSimplifyingSemiSupervised2020}
Kihyuk Sohn, David Berthelot, Chun-Liang Li, Zizhao Zhang, Nicholas Carlini,
  Ekin~D. Cubuk, Alex Kurakin, Han Zhang, and Colin Raffel.
\newblock {{FixMatch}}: {{Simplifying Semi-Supervised Learning}} with
  {{Consistency}} and {{Confidence}}.
\newblock {\em arXiv:2001.07685 [cs, stat]}, November 2020.

\bibitem{zhouInstantTeachingEndtoEndSemiSupervised2021}
Qiang Zhou, Chaohui Yu, Zhibin Wang, Qi~Qian, and Hao Li.
\newblock Instant-{{Teaching}}: {{An End-to-End Semi-Supervised Object
  Detection Framework}}.
\newblock In {\em 2021 {{IEEE}}/{{CVF Conference}} on {{Computer Vision}} and
  {{Pattern Recognition}} ({{CVPR}})}, pages 4079--4088, {Nashville, TN, USA},
  June 2021. {IEEE}.

\bibitem{mittalSemiSupervisedSemanticSegmentation2021}
Sudhanshu Mittal, Maxim Tatarchenko, and Thomas Brox.
\newblock Semi-{{Supervised Semantic Segmentation With High-}} and {{Low-Level
  Consistency}}.
\newblock {\em IEEE Transactions on Pattern Analysis and Machine Intelligence},
  43(4):1369--1379, April 2021.

\bibitem{sohnSimpleSemiSupervisedLearning2020}
Kihyuk Sohn, Zizhao Zhang, Chun-Liang Li, Han Zhang, Chen-Yu Lee, and Tomas
  Pfister.
\newblock A {{Simple Semi-Supervised Learning Framework}} for {{Object
  Detection}}.
\newblock {\em arXiv:2005.04757 [cs]}, December 2020.

\bibitem{zhouSSMDSemiSupervisedMedical2021}
Hong-Yu Zhou, Chengdi Wang, Haofeng Li, Gang Wang, Shu Zhang, Weimin Li, and
  Yizhou Yu.
\newblock {{SSMD}}: {{Semi-Supervised}} medical image detection with adaptive
  consistency and heterogeneous perturbation.
\newblock {\em Medical Image Analysis}, 72:102117, August 2021.

\bibitem{liuSemiSupervisedMedicalImage2020}
Quande Liu, Lequan Yu, Luyang Luo, Qi~Dou, and Pheng~Ann Heng.
\newblock Semi-{{Supervised Medical Image Classification With Relation-Driven
  Self-Ensembling Model}}.
\newblock {\em IEEE Transactions on Medical Imaging}, 39(11):3429--3440, 2020.

\bibitem{huoAutomaticGradingAssessments2022}
Jiayu Huo, Xi~Ouyang, Liping Si, Kai Xuan, Sheng Wang, Weiwu Yao, Ying Liu, Jia
  Xu, Dahong Qian, Zhong Xue, Qian Wang, Dinggang Shen, and Lichi Zhang.
\newblock Automatic {{Grading Assessments}} for {{Knee MRI Cartilage Defects}}
  via {{Self-ensembling Semi-supervised Learning}} with {{Dual-Consistency}}.
\newblock {\em Medical Image Analysis}, page 102508, June 2022.

\bibitem{luoSemisupervisedMedicalImage2021}
Xiangde Luo, Jieneng Chen, Tao Song, and Guotai Wang.
\newblock Semi-supervised {{Medical Image Segmentation}} through {{Dual-task
  Consistency}}.
\newblock {\em Proceedings of the AAAI Conference on Artificial Intelligence},
  35(10):8801--8809, May 2021.

\bibitem{bakaUltrasoundAidedVertebral2017}
Nora Baka, Sieger Leenstra, and Theo {van Walsum}.
\newblock Ultrasound {{Aided Vertebral Level Localization}} for {{Lumbar
  Surgery}}.
\newblock {\em IEEE Transactions on Medical Imaging}, 36(10):2138--2147, 2017.

\bibitem{tranAutomaticDetectionLumbar2010}
Denis Tran and Robert~N. Rohling.
\newblock Automatic {{Detection}} of {{Lumbar Anatomy}} in {{Ultrasound
  Images}} of {{Human Subjects}}.
\newblock {\em IEEE Transactions on Biomedical Engineering}, 57(9):2248--2256,
  September 2010.

\bibitem{bertonSegmentationSpinousProcess2016}
Florian Berton, Farida Cheriet, Marie-Claude Miron, and Catherine Laporte.
\newblock Segmentation of the spinous process and its acoustic shadow in
  vertebral ultrasound images.
\newblock {\em Computers in Biology and Medicine}, 72, March 2016.

\bibitem{liaoJointVertebraeIdentification2018}
Haofu Liao, Addisu Mesfin, and Jiebo Luo.
\newblock Joint {{Vertebrae Identification}} and {{Localization}} in {{Spinal
  CT Images}} by {{Combining Short-}} and {{Long-Range Contextual
  Information}}.
\newblock {\em IEEE Transactions on Medical Imaging}, 37(5):1266--1275, 2018.

\bibitem{yangAutomaticVertebraLabeling2017}
Dong Yang, Tao Xiong, Daguang Xu, Qiangui Huang, David Liu, S.~Kevin Zhou,
  Zhoubing Xu, JinHyeong Park, Mingqing Chen, Trac~D. Tran, Sang~Peter Chin,
  Dimitris Metaxas, and Dorin Comaniciu.
\newblock Automatic {{Vertebra Labeling}} in {{Large-Scale 3D CT}} using {{Deep
  Image-to-Image Network}} with {{Message Passing}} and {{Sparsity
  Regularization}}.
\newblock {\em arXiv:1705.05998 [cs]}, May 2017.

\bibitem{chenVertebraeIdentificationLocalization2020}
Yizhi Chen, Yunhe Gao, Kang Li, Liang Zhao, and Jun Zhao.
\newblock Vertebrae {{Identification}} and {{Localization Utilizing Fully
  Convolutional Networks}} and a {{Hidden Markov Model}}.
\newblock {\em IEEE Transactions on Medical Imaging}, 39(2):387--399, February
  2020.

\bibitem{wuAutomaticLandmarkEstimation2017}
Hongbo Wu, Chris Bailey, Parham Rasoulinejad, and Shuo Li.
\newblock Automatic {{Landmark Estimation}} for {{Adolescent Idiopathic
  Scoliosis Assessment Using BoostNet}}.
\newblock In Maxime Descoteaux, Lena {Maier-Hein}, Alfred Franz, Pierre Jannin,
  D.~Louis Collins, and Simon Duchesne, editors, {\em Medical {{Image
  Computing}} and {{Computer Assisted Intervention}} - {{MICCAI}} 2017}, volume
  10433, pages 127--135. {Springer International Publishing}, {Cham}, 2017.

\bibitem{sunDirectEstimationSpinal2017}
Haoliang Sun, Xiantong Zhen, Chris Bailey, Parham Rasoulinejad, Yilong Yin, and
  Shuo Li.
\newblock Direct {{Estimation}} of {{Spinal Cobb Angles}} by {{Structured
  Multi-output Regression}}.
\newblock In Marc Niethammer, Martin Styner, Stephen Aylward, Hongtu Zhu, Ipek
  Oguz, Pew-Thian Yap, and Dinggang Shen, editors, {\em Information
  {{Processing}} in {{Medical Imaging}}}, Lecture {{Notes}} in {{Computer
  Science}}, pages 529--540, {Cham}, 2017. {Springer International Publishing}.

\bibitem{zhangLearningbasedCoronalSpine2021}
T.~Zhang, Y.~Li, J.~P.~Y. Cheung, S.~Dokos, and K.-Y.~K. Wong.
\newblock Learning-based coronal spine alignment prediction using
  smartphone-acquired scoliosis radiograph images.
\newblock {\em IEEE Access}, pages 1--1, 2021.

\bibitem{wuAutomatedComprehensiveAdolescent2018}
Hongbo Wu, Chris Bailey, Parham Rasoulinejad, and Shuo Li.
\newblock Automated comprehensive {{Adolescent Idiopathic Scoliosis}}
  assessment using {{MVC-Net}}.
\newblock {\em Medical Image Analysis}, 48:1--11, August 2018.

\bibitem{newellStackedHourglassNetworks2016}
Alejandro Newell, Kaiyu Yang, and Jia Deng.
\newblock Stacked {{Hourglass Networks}} for {{Human Pose Estimation}}.
\newblock In Bastian Leibe, Jiri Matas, Nicu Sebe, and Max Welling, editors,
  {\em Computer {{Vision}} \textendash{} {{ECCV}} 2016}, Lecture {{Notes}} in
  {{Computer Science}}, pages 483--499, {Cham}, 2016. {Springer International
  Publishing}.

\bibitem{wangDeepHighResolutionRepresentation2021}
Jingdong Wang, Ke~Sun, Tianheng Cheng, Borui Jiang, Chaorui Deng, Yang Zhao,
  Dong Liu, Yadong Mu, Mingkui Tan, Xinggang Wang, Wenyu Liu, and Bin Xiao.
\newblock Deep {{High-Resolution Representation Learning}} for {{Visual
  Recognition}}.
\newblock {\em IEEE Transactions on Pattern Analysis and Machine Intelligence},
  43(10):3349--3364, 2021.

\bibitem{leePseudoLabelSimpleEfficient2013}
Dong-Hyun Lee.
\newblock Pseudo-{{Label}} : {{The Simple}} and {{Efficient Semi-Supervised
  Learning Method}} for {{Deep Neural Networks}}.
\newblock page~6, 2013.

\bibitem{berthelotMixMatchHolisticApproach2019a}
David Berthelot, Nicholas Carlini, Ian Goodfellow, Nicolas Papernot, Avital
  Oliver, and Colin~A Raffel.
\newblock {{MixMatch}}: {{A Holistic Approach}} to {{Semi-Supervised
  Learning}}.
\newblock In {\em Advances in {{Neural Information Processing Systems}}},
  volume~32. {Curran Associates, Inc.}, 2019.

\bibitem{laineTemporalEnsemblingSemiSupervised2017}
Samuli Laine and Timo Aila.
\newblock Temporal {{Ensembling}} for {{Semi-Supervised Learning}}, March 2017.

\bibitem{tarvainenMeanTeachersAre2017}
Antti Tarvainen and Harri Valpola.
\newblock Mean teachers are better role models: {{Weight-averaged}} consistency
  targets improve semi-supervised deep learning results.
\newblock In {\em Advances in {{Neural Information Processing Systems}}},
  volume~30. {Curran Associates, Inc.}, 2017.

\bibitem{wangFocalMixSemiSupervisedLearning2020}
Dong Wang, Yuan Zhang, Kexin Zhang, and Liwei Wang.
\newblock {{FocalMix}}: {{Semi-Supervised Learning}} for {{3D Medical Image
  Detection}}.
\newblock In {\em Proceedings of the {{IEEE}}/{{CVF Conference}} on {{Computer
  Vision}} and {{Pattern Recognition}}}, pages 3951--3960, 2020.

\bibitem{pengCMCNet3DCalf2022}
Yaopeng Peng, Hao Zheng, Lichun Zhang, Milan Sonka, and Danny~Z. Chen.
\newblock {{CMC-Net}}: {{3D}} calf muscle compartment segmentation with sparse
  annotation.
\newblock {\em Medical Image Analysis}, 79:102460, July 2022.

\bibitem{liDeepWeaklysupervisedBreast2022}
Yongshuai Li, Yuan Liu, Lijie Huang, Zhili Wang, and Jianwen Luo.
\newblock Deep weakly-supervised breast tumor segmentation in ultrasound images
  with explicit anatomical constraints.
\newblock {\em Medical Image Analysis}, 76:102315, February 2022.

\bibitem{morrissyMeasurementCobbAngle1990}
R~T Morrissy, G~S Goldsmith, E~C Hall, D~Kehl, and G~H Cowie.
\newblock Measurement of the {{Cobb}} angle on radiographs of patients who have
  scoliosis. {{Evaluation}} of intrinsic error.:.
\newblock {\em The Journal of Bone \& Joint Surgery}, 72(3):320--327, March
  1990.

\bibitem{herzenbergCobbAngleSpinous1990}
John~E. Herzenberg, Nicholas~A. Waanders, Robert~F. Closkey, Albert~B. Schultz,
  and Robert~N. Hensinger.
\newblock Cobb {{Angle Versus Spinous Process Angle}} in {{Adolescent
  Idiopathic Scoliosis The Relationship}} of the {{Anterior}} and {{Posterior
  Deformities}}:.
\newblock {\em Spine}, 15(9):874--879, September 1990.

\bibitem{zengAutomaticDetectionMeasurement2021}
Hong-Ye Zeng, Edmond Lou, Song-Han Ge, Zhi-Chao Liu, and Rui Zheng.
\newblock Automatic {{Detection}} and {{Measurement}} of {{Spinous Process
  Curve}} on {{Clinical Ultrasound Spine Images}}.
\newblock {\em IEEE Transactions on Ultrasonics, Ferroelectrics, and Frequency
  Control}, 68(5):1696--1706, May 2021.

\bibitem{geAutomaticMeasurementSpinous2020}
S.~Ge, H.~Zeng, and R.~Zheng.
\newblock Automatic {{Measurement}} of {{Spinous Process Angles}} on
  {{Ultrasound Spine Images}}.
\newblock In {\em 2020 42nd {{Annual International Conference}} of the {{IEEE
  Engineering}} in {{Medicine Biology Society}} ({{EMBC}})}, pages 2101--2104,
  July 2020.

\bibitem{morrisonCorrelationCobbAngle2015}
Devlin~G. Morrison, Amanda Chan, Doug Hill, Eric~C. Parent, and Edmond H.~M.
  Lou.
\newblock Correlation between {{Cobb}} angle, spinous process angle ({{SPA}})
  and apical vertebrae rotation ({{AVR}}) on posteroanterior radiographs in
  adolescent idiopathic scoliosis ({{AIS}}).
\newblock {\em Eur Spine J}, 24(2):306--312, February 2015.

\end{thebibliography}

\end{document}